\documentstyle[prb,aps]{revtex}
\input BoxedEPS.tex
\SetOzTeXEPSFSpecial
\HideDisplacementBoxes

\begin{document}

%\draft{}
\twocolumn[\hsize\textwidth\columnwidth\hsize
           \csname @twocolumnfalse\endcsname
\title{Pairing interactions and pairing mechanism in high temperature 
copper oxide superconductors} 
\author{Guo-meng Zhao$^{*}$} 
\address{Department of Physics and Astronomy, 
California State University, Los Angeles, CA 90032, USA}

\maketitle
\widetext

\begin{abstract}
The polaron binding energy $E_{p}$ in undoped parent cuprates has 
been 
determined to be about 1.0 eV from the unconventional oxygen-isotope 
effect on the 
antiferromagnetic ordering temperature. The deduced value of $E_{p}$ 
is in quantitative agreement with that estimated from independent 
optical data and that estimated theoretically from the measured 
dielectric constants. The 
substantial oxygen-isotope effect on the in-plane 
supercarrier mass observed in optimally doped cuprates suggests that 
polarons 
are bound into the Cooper pairs.  We 
also identify the phonon modes that are strongly coupled to 
conduction electrons from the angle-resolved photoemission 
spectroscopy, 
tunneling spectra, and optical data. We consistently show 
that there is a very strong electron-phonon coupling feature at a phonon energy of 
about 20 meV along the antinodal direction and that this 
coupling becomes weaker towards the diagonal 
direction. We further show  that 
high-temperature superconductivity in cuprates is caused by strong 
electron-phonon coupling, polaronic effect, and significant coupling 
with 2 eV Cu-O charge transfer fluctuation.
 
\end{abstract}
\vspace{0.5cm}
\narrowtext
%\newpage
]
\section{Introduction}
Developing the microscopic theory for high-$T_c$ superconductivity is 
one of the most challenging problems in condensed matter physics.  
Eighteen years after the discovery of the high-$T_c$ cuprate 
superconductors by 
Bednorz and M\"uller \cite{KAM86}, there have been no microscopic 
theories
that can describe the physics
of high-$T_c$ superconductors completely and unambiguously. Due to 
the 
high $T_c$ values and the observation of a small oxygen-isotope 
effect on $T_{c}$ in a 90 K cuprate superconductor 
YBa$_{2}$Cu$_{3}$O$_{7-y}$ (YBCO) 
\cite{Cava,Batlogg,Bourne87}, many theorists 
believe that the electron-phonon interaction is not important in 
bringing about high-$T_c$ superconductivity. Most physicists have 
thus turned their 
minds towards alternative pairing interactions of purely electronic
origin.

On the other hand, there is overwhelming evidence that 
electron-phonon coupling is very strong 
in the cuprate superconductors 
\cite{ZhaoAF,ZhaoYBCO,ZhaoLSCO,ZhaoNature97,ZhaoJPCM,McQueeney,Lanzara,SG,Mis,HoferPRL,Zhaoreview1,Zhaoreview2,Zhaoisotope,Lanzara01,Keller1,Zhou04}. 
In particular, various unconventional oxygen-isotope 
effects Zhao and his 
coworkers have observed since 1994 clearly indicate that the 
electron-phonon interactions are so strong that polarons/bipolarons 
are formed in doped cuprates 
\cite{ZhaoAF,ZhaoYBCO,ZhaoLSCO,ZhaoNature97,ZhaoJPCM,Lanzara,SG,HoferPRL,Zhaoreview1,Zhaoreview2,Zhaoisotope,Keller1} and manganites 
\cite{ZhaoNature96,Zhaobook}, 
in agreement with a theory of 
high-temperature superconductivity \cite{ale} and the original 
motivation for the discovery of high-temperature superconductivity 
\cite{KAM86}. However, such clear 
experimental evidence for strong electron-phonon interactions from 
the unconventional isotope effects has been generally ignored.  In 
the 2001 Nature paper 
\cite{Lanzara01}, Lanzara 
{\em et al.} appear to provide evidence for strong coupling between 
doped 
holes and the 70 meV half-breathing phonon mode from angle-resolved 
photoemission 
spectroscopy (ARPES).  They further show that this 70 meV phonon mode 
can lead 
to d-wave pairing symmetry and is mainly responsible for 
high-temperature superconductivity \cite{Shen}. Very recently, 
Devereaux {\em et al.} \cite{Devereaux04} have proposed that the 40 
meV $B_{1g}$ 
phonon mode rather than the 70 meV half-breathing phonon mode is 
responsible for 
d-wave high-temperature superconductivity.  This pairing mechanism 
contradicts 
the very recent ARPES data, which show that 
multiple phonon modes at 27 meV, 45 meV, 61 meV, and 75 meV are 
strongly coupled to doped holes in deeply underdoped 
La$_{2-x}$Sr$_{x}$CuO$_{4}$ \cite{Zhou04}. The strong coupling to the 
multiple 
phonon modes is not in favor of d-wave gap symmetry but may support a 
general s-wave gap symmetry \cite{Zhaosymmetry,Brandow}.

Here we determine the polaron binding energy $E_{p}$ for undoped 
parent cuprates 
from the unconventional oxygen-isotope effect on the 
antiferromagnetic ordering temperature (T$_{N}$) \cite{ZhaoAF}. The 
determined value ($\sim$1.0 eV) of $E_{p}$ 
is in quantitative agreement with that estimated from independent 
optical data \cite{Bi} and that estimated theoretically from the 
measured 
dielectric constants \cite{Alex99}. The 
substantial oxygen-isotope effect on the in-plane 
supercarrier mass observed in optimally doped cuprates 
\cite{ZhaoYBCO,Zhaoisotope,Keller1} indicates that polarons 
are bound into the Cooper pairs.  We 
also identify the phonon modes that are strongly coupled to 
conduction electrons from the angle-resolved photoemission 
spectroscopy, tunneling spectra, and optical data. We consistently show 
that there is a very strong electron-phonon coupling feature at a phonon energy of 
about 20 meV along the antinodal direction and that this 
coupling becomes weaker towards the diagonal 
direction. We further show  that 
high-temperature superconductivity in cuprates is caused by strong 
electron-phonon coupling, polaronic effect, and significant coupling 
with 2 eV Cu-O charge transfer fluctuation.

\section{Oxygen-isotope effect on $T_{N}$ in La$_{2}$CuO$_{4}$}

The antiferromagnetic order (AF) observed in the parent insulating 
compounds like La$_{2}$CuO$_{4}$ signals a strong electron-electron 
Coulomb correlation.  On the other hand, if there is a very strong 
electron-phonon coupling such that the Migdal adiabatic approximation 
breaks down, one might expect that the 
antiferromagnetic exchange energy should depend on the isotope mass.  
Following this simple argument, Zhao and his co-workers initiated 
studies of the 
oxygen isotope effect on the AF ordering temperature in several 
parent 
compounds in 1992. A noticeable oxygen-isotope shift of $T_{N}$ was 
consistently observed in La$_{2}$CuO$_{4}$ \cite{ZhaoAF}.

\begin{figure}[htb]
\ForceWidth{7cm}
	\centerline{\BoxedEPSF{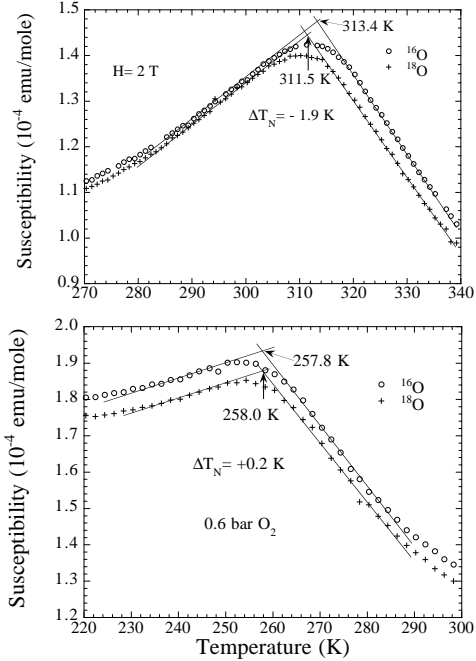}}
\caption[~]{The temperature dependence of the 
susceptibility for the $^{16}$O and $^{18}$O samples of the undoped 
La$_{2}$CuO$_{4}$ (upper panel), and of the oxygen-doped 
La$_{2}$CuO$_{4+y}$ (lower panel).  After 
\cite{ZhaoAF}.}
\label{IF11}
\end{figure}

Fig.~\ref{IF11} shows the temperature dependence of the 
susceptibility for the $^{16}$O and $^{18}$O samples of undoped 
La$_{2}$CuO$_{4}$ (upper panel), and of oxygen doped 
La$_{2}$CuO$_{4+y}$ (lower panel). 
One can see that the AF ordering temperature $T_{N}$ for the $^{18}$O 
sample is lower than the $^{16}$O sample by about 1.9 K in the case 
of the undoped samples. For the oxygen-doped samples, there is a 
negligible isotope effect.

It is known that the antiferromagnetic properties of 
La$_{2}$CuO$_{4+y}$ can be well understood within 
mean-field theory which leads to a $T_{N}$ formula 
\cite{Thio}:
\begin{equation}\label{Ie5}
k_{B}T_{N} = J' [\xi(T_{N})/a]^{2},
\end{equation}
where $J'$ is the interlayer coupling energy, $\xi (T_{N})$ is the 
in-plane AF correlation length at $T_{N}$ with 
$\xi (T_{N}) \propto \exp (J/T_{N})$ for $y$ = 0 ($J$ is the in-plane 
exchange energy). When $T_{N}$ is 
reduced to about 250 K by oxygen doping, a mesoscopic phase 
separation has taken place so that $\xi (T_{N})$ = $L$ 
(Ref.~\cite{Cho}), where $L$ is 
the size of the antiferromagnetically correlated clusters, and 
depends 
only on the extra oxygen content $y$. In this case, we have $T_{N} = 
J'(L/a)^{2}$. Since $L$ is independent of the isotope mass, a 
negligible 
isotope shift of $T_{N}$ in the oxygen-doped La$_{2}$CuO$_{4+y}$ 
suggests that 
$J'$ is independent of the isotope mass. Then we easily find for 
undoped compounds
\begin{equation}\label{Ie6}                                       
\Delta T_{N}/T_{N} = (\Delta J/J)\frac{B}{1+B},
\end{equation}
where $B = 2J/T_{N} \simeq$ 10. From the measured isotope 
shift of $T_{N}$ for the undoped samples, we obtain $\Delta J/J$ 
$\simeq -0.6\%$.

Recently, Eremin {\em et al.} \cite{Eremin} have considered strong 
electron-phonon 
coupling within a three-band Hubbard model.  They showed 
that the antiferromagnetic exchange energy $J$ depends on the polaron 
binding energy $E_{p}^{O}$ due to oxygen vibrations, on the polaron 
binding energy $E_{p}^{Cu}$ due to copper vibrations, and on their 
respective vibration frequencies $\omega_{O}$ and $\omega_{Cu}$. At 
low temperatures, $J$ is given by \cite{Eremin}
\begin{equation}
J = J_{\circ}(1 + 
\frac{3E_{p}^{O}\hbar\omega_{O}}{\Delta_{pd}^{2}}+\frac{3E_{p}^{Cu}\hbar\omega_{Cu}}{\Delta_{pd}^{2}}).
\end{equation}

Here $\Delta_{pd}$ is the charge-transfer gap, which is measured to 
be about 1.5 eV in undoped cuprates. The oxygen-isotope effect on $J$ 
can be readily deduced from Eq.~3:
\begin{equation}\label{Ie8}
\frac{\Delta J}{J} = 
(\frac{3E_{p}^{O}\hbar\omega_{O}}{\Delta_{pd}^{2}})(\frac{\Delta\omega_{O}}{\omega_{O}}).
\end{equation}

Substituting the unbiased parameters $\hbar\omega_{O}$ = 0.075 eV, 
$\Delta J/J$ $\simeq -0.6\%$, 
$\Delta_{pd}$ = 1.5 eV, and $\Delta\omega_{O}/\omega_{O}$ = 6.0$\%$ 
into Eq.~\ref{Ie8}, we find that $E_{p}^{O}$ = 1.0 eV. The total 
polaron binding energy should be larger than 1.0 eV since 
$E_{p}^{Cu}$ 
should not be zero. The parameter-free estimate of the polaron 
binding 
energy due to the long-range Fr\"ohlich-type electron-phonon 
interaction 
has been made for many oxides including cuprates and manganites 
\cite{Alex99}. The 
total polaron binding energy  for La$_{2}$CuO$_{4}$ was estimated to 
be 
about 1 eV (Ref.~\cite{Alex99}), in excellent agreement with the value deduced above from 
the isotope effect. The polaron binding energy can be also estimated from 
optical data where the energy of the mid-infrared peak $E_{m}$ in the 
optical 
conductivity is equal to 2$\gamma E_{p}$ (Ref.\cite{Alex99}), where 
$\gamma$  is  0.2$-$0.3 (Ref.\cite{Alex99}).  The peak position 
$E_{m}$ was 
found to be  about 0.6 eV for 
La$_{1.98}$Sr$_{0.02}$CuO$_{4}$ (Ref.\cite{Bi}), implying that 
$E_{p}$ = 1.0$-$1.5 eV. 
This is in quantitative agreement with the value estimated from the 
isotope effect. These results thus consistently suggest that the 
polaron binding energy of undoped La$_{2}$CuO$_{4}$ is about 1 eV. 
Doping will reduce the value of $E_{p}$ and thus $E_{m}$ due to screening of charged carriers. 
The optical conductivity data indeed show that $E_{m}$ = 0.44 eV and 
0.12 eV for $x$ = 0.06 and 0.15, respectively \cite{Bi}.

One may argue that the mid-infrared peak could arise from magnetic 
excitations. A sharp peak feature at about 0.35 eV was seen in the 
optical conductivity of the undoped  YBa$_{2}$Cu$_{3}$O$_{6}$ 
(Ref.~\cite{Grun}). This 
feature was also seen in other undoped cuprates \cite{Perkins,Choi}, and can be well 
explained by the phonon assisted two magnon excitation \cite{Choi}. However, this 
sharp feature is very different from a broad peak at about 0.6 eV in lightly 
doped La$_{1.98}$Sr$_{0.02}$CuO$_{4}$. In particular, the 
maximum conductivity for the two magnon peak is about 
1 ($\Omega^{-1}$cm$^{-1}$) \cite{Grun,Perkins,Choi}, which is 
over two orders of magnitude lower than that for the broad peak in 
La$_{1.98}$Sr$_{0.02}$CuO$_{4}$ (Ref.~\cite{Uchida}). Therefore, the broad peak in doped 
La$_{1.98}$Sr$_{0.02}$CuO$_{4}$ cannot have the same origin as the 
sharp peak in the undoped system.

\section{Oxygen-isotope effect on the in-plane supercarrier mass}
 
 One of the most remarkable oxygen-isotope effects we have observed 
is 
 the oxygen-isotope effect on the penetration depth 
\cite{ZhaoYBCO,ZhaoLSCO,ZhaoNature97,ZhaoJPCM,HoferPRL,Zhaoreview1,Zhaoreview2,Zhaoisotope,Keller1}. 
We made the first 
 observation of this effect in optimally doped 
 YBa$_{2}$Cu$_{3}$O$_{6.93}$ in 1994 \cite{ZhaoYBCO}. By precisely 
measuring the 
 diamagnetic signals for the $^{16}$O and $^{18}$O samples, we were 
 able to deduce the oxygen-isotope effects on the penetration depth 
 $\lambda(0)$ and on the supercarrier density $n_{s}$. It turns out 
 that $\Delta n_{s}$ $\simeq$ 0, and $\Delta\lambda(0)/\lambda(0)$ = 
 3.2 $\%$ (Ref.\cite{ZhaoYBCO}). These isotope effects thus suggest 
that the effective 
 supercarrier mass depends on the oxygen-isotope mass.
 
\begin{figure}[htb]
 %\vspace{2cm}
    \ForceWidth{7cm}
	\centerline{\BoxedEPSF{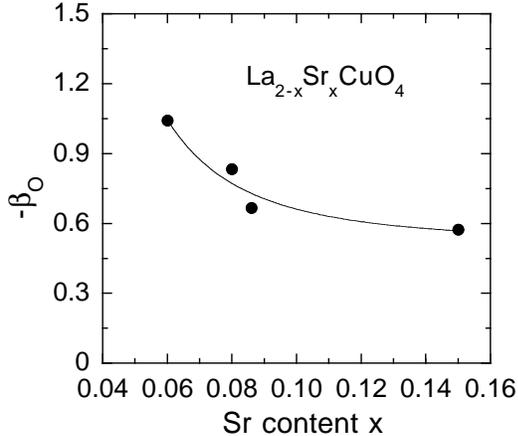}}
	\vspace{1cm}
	\caption[~]{The doping dependence of the exponent ($\beta_{O}$) of the 
oxygen-isotope effect on the in-plane supercarrier mass in  
La$_{2-x}$Sr$_{x}$CuO$_{4}$. The exponent is defined as $\beta_{O}= 
-d\ln m^{**}_{ab}/d\ln M_{O}$. The data are from 
Ref.~\cite{HoferPRL,Zhaoreview2,Zhaoisotope}.}
	\protect\label{OEmass}
\end{figure}

 In fact, for highly anisotropic materials, the observed isotope 
 effect on the angle-averaged $\lambda(0)$ is the same as the isotope 
 effect on the in-plane penetration depth $\lambda_{ab}(0)$. From the 
 magnetic data for YBa$_{2}$Cu$_{3}$O$_{6.93}$, 
 La$_{1.85}$Sr$_{0.15}$CuO$_{4}$, and 
 Bi$_{1.6}$Pb$_{0.4}$Sr$_{2}$Ca$_{2}$Cu$_{3}$O$_{10+y}$, we found 
 that $\Delta\lambda_{ab} 
(0)/\lambda_{ab} (0)$ = 3.2$\pm$0.7$\%$ for the three optimally doped 
cuprates \cite{Zhaoisotope}. Several 
independent  experiments have 
consistently shown that the carrier densities of the two isotope 
samples are the same within 0.0004 per unit cell 
\cite{ZhaoNature97,ZhaoJPCM,Zhaoreview1}. Therefore, we can safely 
conclude that the observed oxygen-isotope effect on the in-plane 
penetration depth is caused only by the isotope dependence of the 
in-plane supercarrier mass $m^{**}_{ab}$.  Recently, direct 
measurements of the 
in-plane penetration 
depth by low energy muon-spin-relaxation (LE$\mu$SR) technique 
\cite{Keller1} have confirmed 
our earlier isotope-effect results. It was found that \cite{Keller1} 
$\Delta\lambda_{ab} 
(0)/\lambda_{ab} (0)$ = 2.8$\pm$1.0$\%$.  It is remarkable that the 
isotope effect obtained from the most advanced technology 
(LE$\mu$SR) \cite{Keller1}
is the same as that deduced from simple magnetic measurements 
\cite{ZhaoYBCO,Zhaoisotope,Keller1}.

Fig.~\ref{OEmass} shows the doping dependence of the exponent 
($\beta_{O}$) of the 
oxygen-isotope effect on the in-plane supercarrier mass in  
La$_{2-x}$Sr$_{x}$CuO$_{4}$. Here the exponent is defined as 
$\beta_{O}= 
-d\ln m^{**}_{ab}/d\ln M_{O}$ (where $M_{O}$ is the oxygen mass). It 
is apparent that the exponent 
increases with decreasing doping, in agreement with the fact that 
doping reduces electron-phonon coupling due to screening.  The large 
oxygen-isotope effect on the in-plane supercarrier 
mass cannot be explained within the conventional phonon-mediated 
pairing mechanism 
where the effective mass of supercarriers is independent of the 
isotope mass \cite{CarbotteRev}. In particular, the substantial 
oxygen-isotope effect 
on $m^{**}_{ab}$ in optimally doped cuprates indicates that the 
polaronic 
effect is not vanished in the optimal doping regime where the 
BCS-like superconducting transition occurs.  This suggests that 
polaronic carriers may be bound into the Cooper pairs in optimally 
doped and overdoped cuprates, in agreement with theory 
\cite{Alex83,ale,Alex03}.

\section{Strong electron-phonon coupling features along the diagonal 
direction}

In conventional superconductors, strong electron-phonon coupling 
features 
can be identified from single-particle tunneling spectra. For 
high-$T_{c}$ cuprates, high-quality tunneling spectra are difficult 
to obtain because of a short coherence length. Moreover, due to a 
strong 
gap anisotropy, the energies of the strong coupling features will 
depend on 
the tunneling directions. Only if one can make a directional 
tunneling, one may be able to accurately identify the electron-phonon coupling 
features from the tunneling spectrum. On the other hand, the 
observation of the electron self-energy
renormalization effect in the form of a ``kink'' in the band
dispersion may
reveal coupling of electrons with phonon modes. The ``kink'' feature 
at an energy of about 65 meV has been seen in the band dispersion of 
various 
cuprate superconductors along the diagonal (``nodal'') direction 
\cite{Lanzara01}. From 
the measured dispersion, one can extract the real part of the electron 
self-energy that contains information about coupling of electrons 
with collective boson modes. The remarkable progress in the ARPES 
experiments is that the fine electron-phonon coupling structures
have been revealed in the high-resolution ARPES data of a Be surface 
\cite{Shi}. Very recently, such 
fine coupling structures have been also seen in the raw data of 
the electron self-energy 
of deeply underdoped La$_{2-x}$Sr$_{x}$CuO$_{4}$ along the diagonal 
direction \cite{Zhou04}. Using the maximum entropy method (MEM) 
procedure, they are 
able to extract the electron-phonon spectral density 
$\alpha^{2}F(\omega)$ that contains coupling features at 27 meV, 45 
meV, 61 meV and 75 meV. The energies of these coupling features are 
one-to-one correspondences to the phonon energies measured by 
inelastic neutron scattering \cite{Zhou04}. These beautiful ARPES data and exclusive data 
analysis \cite{Zhou04} clearly indicate that the phonons rather than 
the magnetic collective mode are responsible for the electron self-energy effect.

\begin{figure}[htb]
%\vspace{2cm}
    \ForceWidth{7cm}
	\centerline{\BoxedEPSF{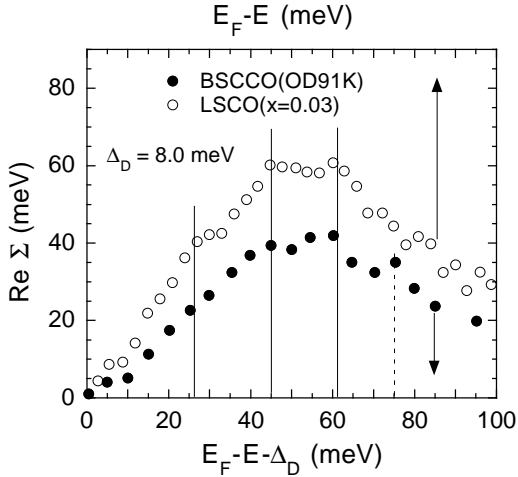}}
	\vspace{0.5cm}
	\caption[~]{The real part of the electron self-energy along the diagonal direction for a slightly overdoped BSCCO with 
$T_{c}$ = 
91 K (OD91K) \cite{Johnson} and for La$_{2-x}$Sr$_{x}$CuO$_{4}$  with 
$x$ = 0.03 (Ref.~\cite{Zhou04}).  The 
energy scale for BSCCO is shifted down by $\Delta_{D}$ = 8.0 meV. The 
solid vertical lines at 27 meV, 45 meV, 
and 61 meV mark the 
energies of the pronounced phonon peaks in the electron-phonon 
spectral 
density $\alpha^{2}F(\omega)$ of La$_{1.97}$Sr$_{0.03}$CuO$_{4}$, which 
is determined from the MEM procedure 
\cite{Zhou04}. The dashed vertical line indicates the energy of an 
extra pronounced  phonon peak (75 meV) in the 
superconducting LSCO with $x$ = 0.07. It is interesting that the coupling feature 
at 27 meV in LSCO appears to shift to a lower energy of about 20 meV in BSCCO.}
	\protect\label{ARPES1}
\end{figure}

Here we will demonstrate that the fine coupling structures also 
appear in 
the earlier high-resolution ARPES data of a slightly overdoped BSCCO 
with $T_{c}$ = 
91 K (OD91K) \cite{Johnson}. Fig.~\ref{ARPES1} shows the real part of 
the electron self-energy 
along the diagonal direction for the  OD91K sample at 70 K. In the same 
figure, we also plot the real part of the electron self-energy 
along the diagonal direction for the nonsuperconducting La$_{2-x}$Sr$_{x}$CuO$_{4}$ (LSCO) 
with 
$x$=0.03. We can clearly see the fine structures in the raw 
data of both LSCO and BSCCO. The solid vertical lines mark the 
energies of the pronounced phonon peaks (27 meV, 45 
meV, 61 meV) in the electron-phonon 
spectral 
density 
$\alpha^{2}F(\omega)$ of the nonsuperconducting 
La$_{1.97}$Sr$_{0.03}$CuO$_{4}$, which is determined from the MEM procedure 
\cite{Zhou04}. The 
dashed vertical line indicates the energy of an extra pronounced  
phonon 
peak (75 meV) in the 
superconducting LSCO with $x$ = 0.07. In order for three pronounced fine 
structures (45 meV, 61 meV, and 75 meV)
in the self-energy of BSCCO to be aligned with those for LSCO, the 
energy scale for BSCCO has to be shifted down by $\Delta_{D}$ = 8.0 
meV. This suggests that the superconducting gap along the diagonal 
direction is about 8 meV at 70 K. Using the BCS temperature 
dependence of the gap, one finds $\Delta_{D}$ = 
10 meV at zero temperature.

The finite superconducting gap of about 10 meV along the diagonal 
direction is consistent with a general s-wave gap symmetry (s + g wave) with 
eight line nodes (g $>$ s) \cite{Zhaosymmetry}. This gap symmetry has double gap 
features at $\Delta_{M}$ and $\Delta_{D}$ in the 
superconducting density of states \cite{Zhaosymmetry}, which can be seen in the 
single-particle tunneling spectra and in the Andreev reflection spectra. 
Various break-junction spectra suggest that $\Delta_{D}$ = 9.5 meV 
and $\Delta_{M}$ = 26 meV in 
an overdoped BSCCO with $T_{c}$ = 89 K (Ref.~\cite{Zhaosymmetry}), 
$\Delta_{D}$ = 12$\pm$1 meV and $\Delta_{M}$ = 24$\pm$1 meV in an overdoped BSCCO with $T_{c}$ = 86 K 
(Ref.~\cite{Bus}), $\Delta_{D}$ = 
7.5-9.0 meV and $\Delta_{M}$ = 15-18 meV in heavily overdoped BSCCOs with $T_{c}$ = 62 K 
(Ref.~\cite{DeWilde,Ozyuzer}). The Andreev reflection spectrum also 
indicates that $\Delta_{D}$ = 13 meV in 
an overdoped BSCCO with $T_{c}$ = 85 K (Ref.~\cite{Sinha}), while the 
other Andreev reflection spectrum shows a gap feature at $\Delta_{M}$ = 25 
meV (Ref.~\cite{Alff}).  Further, the 
Raman data 
\cite{Hewitt} clearly indicate s + g wave gap symmetry in a heavily 
overdoped BSCCO with $T_{c}$ = 55 K. The Raman 
intensities in both $B_{1g}$ and $B_{2g}$ symmetries increase 
linearly with energy up to 
1.3$\Delta_{M}$ and can be extrapolated to nearly zero values at zero energy. 
The linear energy dependence of the Raman intensity up to 1.3$\Delta_{M}$ 
in the $B_{1g}$ symmetry could be  
consistent with either clean s + g wave superconductivity 
\cite{Sacuto} or very dirty 
d-wave superconductivity with $\sqrt{\Gamma\Delta_{M}(0)}$ $\simeq$ 
1.3$\Delta_{M}$ (where $\Gamma$ is the impurity scattering rate) 
\cite{Dev}. Very dirty d-wave superconductivity would 
give a significant residual intensity at zero energy \cite{Dev}, in disagreement with 
experiment \cite{Hewitt}. Very dirty d-wave superconductivity would also give 
a $T^{2}$ 
dependence of the in-plane penetration depth below $T^{*}$ = 
0.83$\sqrt{\Gamma\Delta_{M}(0)}$ $\simeq$ 1.1$\Delta_{M}$ $\simeq$ 180 K 
(Ref.~\cite{Peter}), in contradiction with 
the observed linear-$T$ dependence below 10 K \cite{Coch}. In 
Ref.~\cite{Zhaosymmetry}, the author has consistently   
explained all the relevant experiments in terms of the s + g wave gap 
symmetry and also shown that some phase sensitive experiments apparently 
supporting a d-wave 
order parameter symmetry do not contradict the s + g wave gap 
symmetry. 
 
It is interesting to note that the coupling feature at 75 meV is 
invisible in the deeply underdoped LSCO ($x$ = 0.03), but becomes 
pronounced in the superconducting LSCO ($x$= 0.07) \cite{Zhou04} and in BSCCO 
(OD91K). This is consistent 
with the neutron experiments that clearly demonstrate that the 
coupling to the 75 meV half-breathing mode increases with increasing 
doping \cite{McQueeney}. Further, the coupling feature at 27 meV in 
LSCO appears to shift to a lower energy of about 20 meV in BSCCO (see 
Fig.~\ref{ARPES1}).

\section{Strong electron-phonon coupling features along the antinodal 
direction}

\begin{figure}[htb]
%\vspace{2cm}
   \ForceWidth{7cm}
	\centerline{\BoxedEPSF{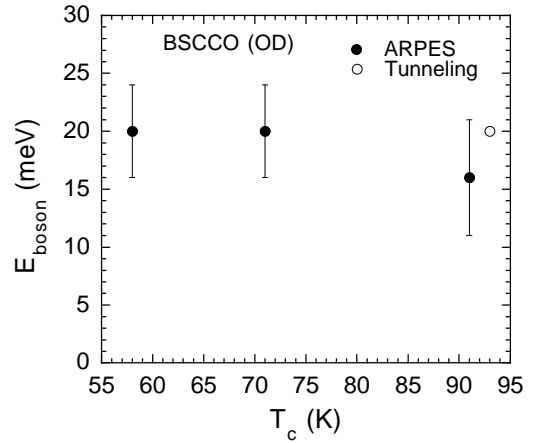}}
	\vspace{0.5cm}
	\caption[~]{The boson energy as a function of $T_{c}$ for 
several overdoped BSCCO. The boson energy extracted from ARPES 
(Ref.~\cite{Gromko}) is calculated according to $E_{boson}$ = $E_{kink} -\Delta_{M}$, where $E_{kink}$ is the 
kink energy in the band dispersion, which is found to be equal to the 
peak energy in
the electron self-energy \cite{Gromko}. One data point (open circle) is from the tunneling 
data (see Fig.~\ref{ARPES3}). }
	\protect\label{ARPES2}
\end{figure}

\begin{figure}[htb]
%\vspace{2cm}
    \ForceWidth{7cm}
	\centerline{\BoxedEPSF{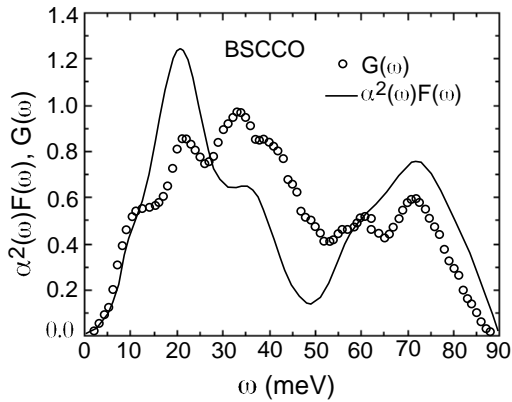}}
	\vspace{0.5cm}
	\caption[~]{The electron-phonon spectral density 
$\alpha^{2} F(\omega)$ for a slightly overdoped 
Bi$_{2}$Sr$_{2}$CaCu$_{2}$O$_{8+y}$ (BSCCO) crystal, which was 
deduced 
from a break-junction spectrum \cite{Gonnelli}. }
	\protect\label{ARPES3}
\end{figure}

The electron self-energy effect along the antinodal direction has 
been 
studied for several BSCCO crystals (OD91K, 
OD71K, and OD58K) \cite{Gromko}.  The kink feature in the band 
dispersion or the peak feature in the electron self-energy 
along the antinodal direction is much stronger than that along 
the diagonal direction. 
This indicates a much stronger electron-boson coupling. One of the 
puzzling issues is that the energies of the boson modes shift 
to a much lower energy (about 20 meV) and are nearly independent of 
doping \cite{Gromko}. Fig.~\ref{ARPES2} shows the 
boson energy as a function of $T_{c}$ for 
several overdoped BSCCO. 
The boson energy from the ARPES data is calculated according to 
$E_{boson}$ = 
$E_{kink} -\Delta_{M}$, where $E_{kink}$ is the 
kink energy in the band dispersion, which is found to be equal to the 
peak energy in
the electron self-energy \cite{Gromko}. The above relation that is 
predicted by theory \cite{Norman} was also used 
by the authors of Ref.~\cite{Gromko} to extract the mode energy.  Since the antinodal gap $\Delta_{M}$ 
is found to be very 
close to the peak energy in the energy 
distribution curve (EDC) \cite{Ding}, one 
can simply take 
$\Delta_{M}$ 
being equal to the EDC peak energy.  It is apparent that the boson 
energy is about 20 
meV for heavily overdoped BSCCOs and about 16 meV for nearly optimally 
doped BSCCO. The strong coupling feature at about 20 meV also agrees with 
the electron-boson spectral density $\alpha^{2}F(\omega)$ deduced 
from a break-junction spectrum of BSCCO (OD93K) \cite{Gonnelli}, as shown in Fig.~\ref{ARPES3}. The 
spectral 
density clearly shows strong coupling features at about 20 meV, 36 
meV, 60 meV, and 72 meV, similar to the features at about 20 meV, 
62 meV, and 75 meV along the diagonal direction (see Fig.~3). Because the 
superconducting gap is very anisotropic, the excellent 
match of the phonon energies obtained from the tunneling spectrum and 
ARPES indicates that the tunneling process of this break-junction 
should be rather directional. Comparing the gap size of 23 meV 
determined from the spectrum \cite{Gonnelli} with the angle dependence 
of the gap \cite{Ding,Zhaosymmetry}, we find that the spectrum may 
mainly probe the superconducting density of states along the directions 
of between 10$^{\circ}$ and 20$^{\circ}$ from the antinodal 
direction. It is interesting that the strong coupling 
feature at 36 meV is only seen in the tunneling spectrum that mainly 
probes the states near the antinodal regime. The 
36 meV phonon mode should be the oxygen buckling mode 
($B_{1g}$) that has been shown to couple more strongly to the states near the antinodal 
direction \cite{Devereaux04}. Such a strong angle dependence of the 
coupling strength may also occur to the other phonon modes such as the 
45 meV mode that has a stronger coupling along the diagonal 
direction. What is more puzzling is that the band dispersion along 
the antinodal direction has a 
single kink feature associated with the 20 meV boson mode.  A very 
likely explanation is that the phonon modes with 
energies higher than 30 meV lie below the bottom of the band continuum 
and thus the kink features for these modes disappear 
\cite{Devereaux04}. Significantly away from the antinodal direction, the energies 
of all the modes are within the band continuum so the kink features for 
these modes will show up.

The much stronger coupling to the 20 meV phonon modes along the 
antinodal direction is rather unusual. This is possible if 
the extended van Hove singularity 
is about 20 meV below the Fermi level and the electron-phonon matrix 
element for the 20 meV phonon modes has a maximum around $\vec{q}$ = 0, where 
$\vec{q}$ is the phonon 
wavevector. The large density of states at the van Hove singularity 
(20 meV below the Fermi level) and strong Fermi surface nesting along 
the antinodal direction greatly enhance the phase space available for 
20 meV small-$\vec{q}$ 
phonons to scatter 
quasiparticles from the states near the antinodal regime to the 
extended saddle points.  The first principle calculation \cite{Pickett91} indeed shows 
that unusual long-range Madelung-like interactions lead to very 
large matrix elements especially for zone center modes ($\vec{q}$ = 
0), which are 
mainly related to vibrations of cations (e.g., La, Sr, Ba, Ca). The phonon 
energies for the vibrations of the cations are between 15 meV to 25 
meV (Ref.~\cite{Pickett91}).

Since the strong coupling feature almost disappears above $T_{c}$, the 
20 meV bosonic mode may be related to the magnetic resonance mode rather 
than the phonons. First, the energy of the bonsic mode is nearly 
independent of $T_{c}$ (see Fig.~\ref{ARPES2}) while the resonance 
energy is proportional 
to $T_{c}$ (Ref.~\cite{He}). This implies that the 20 meV boson is not 
the magnetic resonance mode. Second, 
the disappearance of the coupling feature above $T_{c}$ does not 
necessarily mean that the coupling to the phonon modes is irrelevant. 
Very recent calculation \cite{Devereaux04} shows that the feature of 
coupling to the $B_{1g}$ phonon mode (36 meV) is very weak in the 
normal state. As pointed out by these authors \cite{Devereaux04}, the 
dramatic temperature dependence arises from a 
substantial change in the
electronic occupation distribution and the opening of the
superconducting gap. In the normal state, the phonon
self-energy is a Fermi function at 100 K centered at the
phonon energy, which results in a thermal broadening of 4.4$k_{B}T$ or 38 meV, 
significantly larger than the phonon 
energy (20 meV). At low temperatures and in the 
superconducting state, the phonon self-energy is sharply defined due to 
the step-function-like Fermi function and a singularity
in the superconducting density of states.

Now we further show that a strong coupling feature in 
optical data, which was previously explained 
as due to a strong coupling between electrons and the magnetic 
resonance 
mode \cite{CarbotteNature,Schachinger}, is actually consistent with a 
strong 
electron-phonon coupling at a phonon energy of about 20 meV. It is 
known that the electron-phonon spectral density 
$\alpha^{2}F(\omega)$
can be obtained through inversion of optical data. Marsiglio {\em et 
al.} 
\cite{Marsiglio} introduced a dimensionless 
function $W(\omega)$ which is defined as
the second derivative of the normal state optical scattering rate 
$\tau^{-1}(\omega) = (\Omega_{p}^{2}/4\pi)\Re 
\sigma^{-1}(\omega)$ multiplied 
by frequency $\omega$. Here $\Omega_{p}$ is the bare plasma frequency 
and $\sigma (\omega)$ the normal state
optical conductivity. Specifically,
\begin{equation}\label{Pe1}
W(\omega) = 
\frac{1}{2\pi}\frac{d^{2}}{d\omega^{2}}\frac{\omega}{\tau (\omega)}
\end{equation}
which follows directly from experiment. Marsiglio {\em et
al.} \cite{Marsiglio} made the very important observation that within 
the phonon range 
$W(\omega) \simeq \alpha^{2}F(\omega)$.

\begin{figure}[htb]
%\vspace{0.5cm}
   \ForceWidth{7cm}
	\centerline{\BoxedEPSF{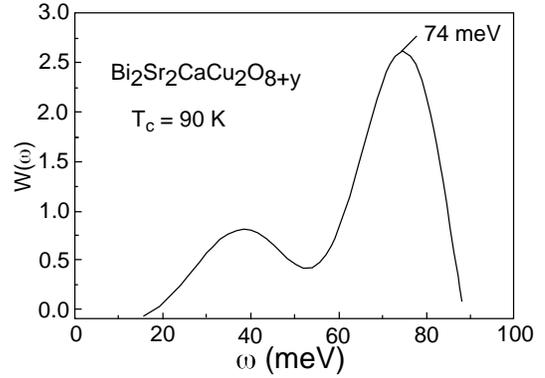}}
	\vspace{0.5cm}
	\caption[~]{The optically determined electron-boson spectral density 
$W(\omega)$ for a slightly overdoped BSCCO crystal with $T_{c}$ = 90 
K. 
After \cite{Schachinger}. }
	\protect\label{ARPES4}
\end{figure}

In the superconducting 
state, a phonon mode that is strongly 
coupled to electrons will appear at an energy of 2$\Delta (\vec{k}) + 
\omega_{ph}$ (where $\omega_{ph}$ is the phonon energy), that is, the 
energies of the phonon structures shift up by the pair-breaking 
energy 2$\Delta (\vec{k})$ \cite{Orenstein}.  Because the 20 meV 
phonon 
modes are much more strongly coupled to the states near the antinodal 
regime 
and because there is a large quasiparticle 
density of states at the maximum gap edge, there must be a maximum at 
2$\Delta_{M} + \omega_{ph}$ in $W(\omega)$. For slightly overdoped 
BSCCO with $T_{c}$ = 90 
K, $\Delta_{M}$ = 26.0(5) meV (Ref.\cite{Kras,Zhaosymmetry}), so we 
should expect a maximum in 
$W(\omega)$ to be at about 72 meV. This is in quantitative agreement 
with the result shown in 
Fig.~\ref{ARPES4}.

Recently, Devereaux {\em et al.} have calculated the electron-phonon 
interactions for the oxygen buckling mode ($B_{1g}$) and the in-plane 
half-breathing mode \cite{Devereaux04}. They find that the 36 meV 
$B_{1g}$ mode 
couples strongly 
to electronic states near the antinodal regime. They use an 
electron-phonon matrix element that is suitable only for 
YBa$_{2}$Cu$_{3}$O$_{7-y}$ where a large buckling distortion 
occurs. For other cuprates, 
the CuO$_{2}$ plane is flat and the buckling effect is negligible. 
Raman data have indeed shown that the coupling constant of the 
$B_{1g}$ mode in BSCCO is more than one order of magnitude smaller 
than that in 
YBCO (Ref.~\cite{Raman1}). Even for YBCO, the coupling constant of 
this mode was deduced to 
be about 0.05 from the Raman data \cite{Raman1}, in agreement with 
the earlier first 
principle calculation \cite{Raman2}. Moreover, if this 40 meV phonon 
were strongly 
coupled to the electronic states near the antinodal regime, one would 
expect a maximum in $W(\omega)$ to occur at about 92 meV in slightly 
overdoped BSCCO with $T_{c}$ = 90 K. This is in disagreement 
with experiment (see Fig.~\ref{ARPES4}).

Previously, the energy of the maximum in $W(\omega)$ was claimed to 
be in quantitative agreement with the theoretical prediction based on 
the strong coupling between electrons and the 
magnetic resonance mode \cite{CarbotteNature,Schachinger}. These 
authors \cite{CarbotteNature,Schachinger} argued that the maximum 
in $W(\omega)$ should occur at about 
$\Delta_{M}+ 
E_{r}$, where $E_{r}$ is the 
magnetic resonance energy. For BSCCO with $T_{c}$ = 90 K, 
$\Delta_{M}$ = 26-28 meV and $E_{r}$ = 
43 meV, so one expects a maximum in $W(\omega)$ to occur at 69-71 meV, 
in reasonable agreement with the experimental result (see Fig.~\ref{ARPES4}). 
Later on, more rigorous 
theoretical approach \cite{Abanov} 
shows that the maximum in $W(\omega)$ should occur at about 
$2\Delta_{M}+ 
E_{r}$ rather than at  $\Delta_{M}+ E_{r}$. Then we should expect a 
maximum in $W(\omega)$ to occur at 96-98 meV, in disagreement with experiment.

If the maximum in $W(\omega)$ would occur at $\Delta_{M} + E_{r}$, the 
electron self-energy determined from the optical data would also have 
a maximum at $\Delta_{M} + E_{r}$. For the optimally doped BSCCO 
(OP96K) with 
$T_{c}$ = 96 K, $\Delta_{M}$ = 37.5 meV (Ref.~\cite{Miy}), and  $E_{r}$ can be estimated 
to be 45 meV using the relation between $E_{r}$ and $T_{c}$ 
(Ref.~\cite{He}). Then the 
maximum in the optically determined electron self-energy would occur at 
82.5 meV. This predicted value is significantly lower than the 
measured one (96 meV) \cite{Hwang}. 

On the other hand, we can 
quantitatively explain the optically determined electron 
self-energy data in terms of the s + g wave gap symmetry and 
electron-phonon coupling. Because the 
superconducting density of states have two sharp maxima at $\Delta_{M}$ 
and $\Delta_{D}$ for the s + g gap symmetry \cite{Zhaosymmetry}, the optically determined electron 
self-energy should have two peak features, one at $2\Delta_{M} + E_{1}$ and 
another at  $2\Delta_{D} + E_{2}$, where $E_{1}$ (= 20 meV) and $E_{2}$ are the 
averaged mode energies along the antinodal and 
diagonal directions, respectively. From 
Fig.~\ref{ARPES1}, we find that $E_{2}$ = 53 meV, which is a simple 
average of the two pronounced peak energies (45 meV and 61 meV) in the electron 
self-energy.  For OP96K, we 
predict two peak features at 95 meV and 73 meV in the optically 
determined electron 
self-energy. The predicted 
73 meV peak feature will show up as a shoulder below the dominant peak 
feature at 95 meV. For an overdoped BSCCO with $T_{c}$ = 82 K, 
$\Delta_{M}$ = 22-25 
meV (Ref.~\cite{Bus,Miy}) and $\Delta_{D}$ $\simeq$ 12 meV (see above), so two 
peak features will 
be located at 64-70 meV and 77 meV, respectively. For an 
overdoped BSCCO with $T_{c}$ = 60 K, $\Delta_{M}$ = 14 
meV and $\Delta_{D}$ = 9 meV (Ref.~\cite{Vob}), so two peak features will 
show up at 48 meV and 71 meV, respectively. All these predicted peak 
features are in quantitative agreement with experiment \cite{Hwang}.

It is interesting to note that the spectral density shown in 
Fig.~\ref{ARPES3} is extracted from a break-junction spectrum that 
has a clear dip feature at an energy of about 47 meV above the 
gap \cite{Gonnelli}. The similar dip features are also seen in 
the ARPES spectra along the antinodal direction 
\cite{Bor}. In fact, the dip features of the 
superconducting density of states occur 
approximately at the valley energies and cut-off energy of the spectral 
density \cite{CarbotteRev,Gonnelli,Um}. For 
example, the cut-off energy of the spectral density for Pb is about 9 
meV (see Fig.~14 of Ref.~\cite{CarbotteRev}) and the dip feature 
also occurs at 9 meV (see Fig.~14 of Ref.~\cite{CarbotteRev}). On the 
other hand, if there is a single sharp peak in the spectral density, 
the dip feature will be slightly above the mode energy \cite{Zasa}. This implies that the 
dip energy measured from the gap is the upper limit of the mode 
energy. Because the peak features in the spectral density can be broadened by the 
strong coupling 
effect and disorder, the dip feature should shift to a higher 
energy towards the underdoping region where the coupling is much 
stronger \cite{Hwang}. In the heavily overdoped 
region, the coupling is weak \cite{Hwang} and the peak width is narrow, so the 
dip energy is close to the mode energy.  This can naturally 
explain why the dip energy of the UD70K sample is about 50 meV (Ref.~\cite{Oz}), while the 
dip energy of the OD62K sample is about 24 meV (Ref.~\cite{Oz}), 
which is slightly greater than the mode energy deduced from the electron self-energy 
above. In contrast, if one assumes that the dip energy measured from 
the gap is equal to the magnetic resonance energy, one cannot 
selfconsistently explain the dip energy of 50 meV in the UD70K 
sample and of 28 meV in the OD87 K sample \cite{Han}. The magnetic resonance 
energies for 
the UD70K and OD87K
samples should be about 30 meV and 40 meV, respectively \cite{He}. 

\section{Strong coupling between electrons and Cu-O charge-transfer 
excitation.}
\begin{figure}[htb]
%\vspace{0.5cm}
   \ForceWidth{7cm}
	\centerline{\BoxedEPSF{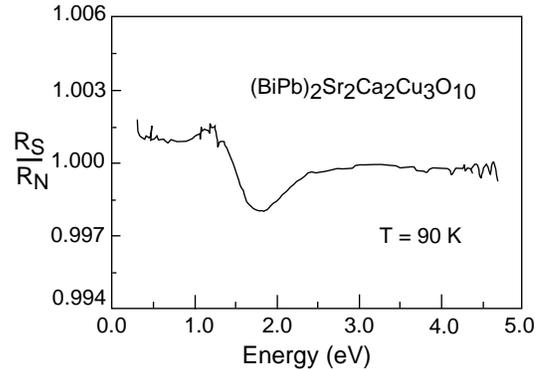}}
	\vspace{0.5cm}
	\caption[~]{The superconducting to normal-state reflectance 
ratio, $R_{s}/R_{N}$, for 
(BiPb)$_{2}$Sr$_{2}$Ca$_{2}$Cu$_{3}$O$_{10}$ with $T_{c}$ = 105 K. 
The figure is reproduced from \cite{Little2}. }
	\protect\label{pla}
\end{figure}
In addition to strong electron-phonon interactions, there is a 
pronounced coupling feature at an energy of about 2 eV in the optical 
reflectance data \cite{Little1,Little2}. In 
Fig.~\ref{pla}, we plot the superconducting to normal-state 
reflectance 
ratio, $R_{s}/R_{N}$ for 
(BiPb)$_{2}$Sr$_{2}$Ca$_{2}$Cu$_{3}$O$_{10}$. It is clear that  
a strong coupling feature appears at about 2 eV. A similar strong 
coupling 
feature was also seen in YBCO, 
Tl$_{2}$Ba$_{2}$Ca$_{2}$Cu$_{3}$O$_{10}$, and 
Tl$_{2}$Ba$_{2}$Ca$_{1}$Cu$_{2}$O$_{8}$ (Ref.~\cite{Little2}).  Both 
the
temperature and energy dependence of the optical structure can be 
well described within
Eliashberg theory with an electron-boson coupling constant of 
0.30-0.35 (Ref.~\cite{Little2}). 
Because the energy scale of this bosonic excitation is similar to the 
Cu-O charge transfer gap, it is likely that this high-energy bosonic 
mode 
corresponds to the Cu-O charge transfer excitation. A recent 
calculation based on 
three-band Hubbard model has indeed shown that the high energy Cu-O 
charge fluctuation can 
lead to a significant attractive interaction between conduction 
electrons and 
that the pairing symmetry is of extended s-wave ($A_{1g}$) 
\cite{Miklos}.  The 
extended s-wave pairing symmetry is consistent with the conclusion 
\cite{Zhaosymmetry}
drawn from the comprehensive data analyses on nearly all the 
experiments that are used to test the gap symmetry.

\section{Pairing mechanism}

In two of our previous papers \cite{Zhaoreview2,Zhaoisotope}, we have 
proposed the pairing mechanism for 
optimally doped and overdoped cuprates. The long-range 
Fr\"ohlich-type electron-phonon 
interaction and the short-range interaction of electrons with 
high-energy 
phonons lead to the formation of polarons. These interactions along 
with the coupling of electrons to the high-energy electronic 
excitations produce a negative value for the effective Coulomb 
pseudopotential $\mu^{*}$. The polarons are bound into the Cooper 
pairs 
due to the negative $\mu^{*}$ and additional attractive interaction 
caused by the retarded electron-phonon interaction with the 20 meV 
phonon 
modes. The problem could 
then be solved within 
Eliashberg equations with an effective electron-phonon spectral 
density 
for the low-energy phonons and a negative Coulomb pseudopotential 
produced by the high-energy phonons and other high-energy bosonic 
excitations of purely electronic origin. Within this simplified 
approach, we are able to consistently explain the observed negligible 
isotope effect on 
$T_{c}$, substantial isotope effect on the supercarrier mass, large 
reduced energy gap, and  high $T_{c}$ value.

In this modified strong-coupling model, the effective 
electron-phonon coupling 
constant $\lambda_{ep}$ for the low-energy phonons is enhanced by 
a factor of $f_{p} = \exp (g^{2})$. Here $g^{2} = A/\omega_{H}$, $A$ 
is a constant, and $\omega_{H}$ is the frequency of the high-energy 
phonon mode \cite{Zhaoisotope}.  The value of $g^{2}$ can be evaluated from the 
mid-infrared optical conductivity which exhibits a maximum at 
$E_{m}$ $\simeq$ 0.12 eV for optimally doped BSCCO \cite{Quijada} and 
LSCO 
\cite{Bi}. With $E_{m}$ = 0.12 eV, 
$\hbar\omega_{H}$ = 
75 meV,  we find $g^{2} = E_{m}/(2\hbar\omega_{H})$ = 0.8, leading to 
$f_{p}$ = 2.2.

From the spectral density shown in Fig.~\ref{ARPES3}, we can 
extract the effective electron-phonon coupling constant  
$\lambda_{ep}$ for the low-energy phonon mode, that is, 
$\lambda_{ep}$ $\simeq$ 2.6. If there were no polaronic mass 
enhancement, the coupling constant 
contributed from the low-energy phonons would be 2.6/$f_{p}$= 
1.2. With $\mu^{*}$ = 0.1 and $\lambda_{ep}$ = 1.2, we calculate 
$T_{c}$ = 18 K according to a $T_{c}$ formula \cite{Kresin}
\begin{equation}\label{Te6}
k_{B}T_{c} = 0.25\hbar\sqrt{<\omega^{2}>}[\exp (2/\lambda_{eff}) - 
1]^{-1/2},
\end{equation}
where
\begin{equation}\label{Te7}
\lambda_{eff} = (\lambda_{ep}-\mu^{*})/[1 + 2\mu^{*} + 
\lambda_{ep}\mu^{*}t(\lambda_{ep})],
\end{equation}
The function $t(\lambda_{ep})$ is plotted in Fig.~2 of 
Ref.\cite{Kresin}. In the present case, $\hbar\sqrt{<\omega^{2}>}$ is 
contributed only from the low-energy phonons and equal to 20 meV. 
Therefore, without the 
polaronic effect, $T_{c}$ would not be higher than 20 K. On the other 
hand, with the polaronic effect, $\lambda_{ep}$ = 2.6 and $\mu^{*}$ 
may 
be close to zero, leading to a $T_{c}$ of about 54 K. Thus the 
polaronic effect enhances $T_{c}$ significantly, but 
electron-phonon coupling alone cannot explain superconductivity above 
100 K in optimally doped cuprates unless the polaronic effect can make 
$\mu^{*}$ $<$ -0.15.  

In order to explain superconductivity above 100 K, it may be essential to 
consider the coupling to the high-energy electronic excitations. One 
of the high-energy excitations is the Cu-O charge transfer excitation 
at about 2 eV, as seen from the optical experiments 
\cite{Little1,Little2}.  Since the high-energy phonon modes couple to 
electrons nonadiabatically, it is likely that the coupling of 
electrons to 
the 2 eV boson mode should also be nonadiabatic. Here we simply ignore 
the nonadiabaticity, and consider two $\delta$-functions in the 
electron-boson spectral density to estimate $T_{c}$. One 
$\delta$-function is at $\hbar\omega_{1}$ = 20 meV with the coupling 
constant $\lambda_{1}$ = 2.6, another at $\hbar\omega_{2}$ = 2100 meV with 
the coupling 
constant $\lambda_{2}$ = 0.3. By solving the s-wave Eliashberg equations
with the above spectral density, we find that $T_{c}$ = 106 K 
for $\mu^{*}$ = 0.1.  In order to obtain $T_{c}$ = 105 K with one $\delta$-function, we 
need the following parameters:  $\lambda_{1}$ = 2.9, $\hbar\omega_{1}$ = 39.5 meV, and 
$\mu^{*}$ = 0.1. If we take $\mu^{*}$ = 0, we need  $\lambda_{1}$ = 2.9 
and $\hbar\omega_{1}$ = 33.6 meV to have $T_{c}$ = 105 K.   Because we have used the
$\delta$-functions, the calculated $T_{c}$ values should be the upper limits. 

\section{Conclusion}

We have determined the polaron binding energy $E_{p}$ for undoped 
parent 
cuprates from the unconventional oxygen-isotope 
effect on the 
antiferromagnetic ordering temperature (T$_{N}$). The deduced value 
(about 1.0 eV) of $E_{p}$ 
is in quantitative agreement with that estimated from independent 
optical data and that estimated theoretically from the measured 
dielectric constants. The polaron binding energy should be large 
enough to overcome the intersite Coulomb interaction to form 
intersite 
bipolarons in deeply underdoped cuprates, in agreement with theory 
and experiment \cite{ale,Zhaoreview2}. The 
substantial oxygen-isotope effect on the in-plane 
supercarrier mass observed in optimally doped cuprates suggests that 
polarons 
are bound into the Cooper pairs.  The bipolaron picture may be 
irrelevant for optimally 
doped cuprates because the superconducting transition is of  
BCS-like. We 
also identify the phonon modes that are strongly coupled to 
conduction electrons from the angle-resolved photoemission 
spectroscopy 
(ARPES), tunneling spectra, and optical data. We consistently show 
that there is a very strong electron-phonon coupling feature at a phonon energy of 
about 20 meV along the antinodal direction and that this 
coupling becomes weaker towards the diagonal 
direction. We further show  that 
high-temperature superconductivity in cuprates is caused by strong 
electron-phonon coupling, polaronic effect, and significant coupling 
with 2 eV Cu-O charge transfer fluctuation. The role of 
the antiferromagnetism in superconductivity appears to be 
insignificant because the antiferromagnetic and superconducting 
phases do not mix, as shown recently by Bozovic and coworkers \cite{Ivan}. 

~\\
~\\
$^{*}$ gzhao2@calstatela.edu

\end{document}